\documentstyle[12pt]{article}
\begin{document}
\begin{titlepage}
\title {Quantum Mechanics
in  Riemannian  Space--Times.\\ I. The Canonical Approach}
\thanks{gr-qc/9812084}
\author {\large E.A.Tagirov \\
 N.N.Bogoliubov Laboratory of Theoretical Physics\\
Joint Institute for Nuclear Research, Dubna, 141980, Russia \\
   e--mail: tagirov@thsun1.jinr.dubna.su }
\date{}
\end{titlepage}
\maketitle

\begin{abstract}
This paper is the first of  two papers devoted to formulation of
quantum mechanics of a particle in a normal geodesic frame of reference
in the general Riemannian space--time. Here  canonical quantization of
geodesic motion in the 1+3--formalism is considered, the result of
which will be compared  in the subsequent paper II with that of the
field--theoretical approach, see also gr-qc/9807030.
The Schr\"odinger representation of quantum-mechanical kinematics and
dynamics is  presented in the general--covariant
form and a physical interpretation of the state vectors and  the
position operators  is discussed.
\end{abstract}

\newcommand{\bgt}{\bigotimes}
\newcommand{\Oc}{O\left(c^{-2(N+1)}\right)}
\newcommand{\im}{\imath}
\newcommand{\Sg}{\Sigma}
\newcommand{\ptl}{\partial}
\newcommand{\Sn}{ Sniatycki}
\newcommand{\Schr}{ Schr\"odinger representation}
\newcommand{\Schp}{ Schr\"odinger picture}
\newcommand{\Sche}{ Schr\"odinger equation}
\newcommand{\eu}{$ E_{1,3} $}
\newcommand{\rif}{ V_{1,3} }
\newcommand{\ri}{$ V_{1,3} $ }
\newcommand{\ov}{\overline}
\newcommand{\stc}{\stackrel{def}{=}}
\newcommand{\h}{\hbar}
\newcommand{\vp}{\varphi}
\newcommand{\1}{{\bf\hat 1}}
\newcommand{\bfc}{{\bf C}}
\newcommand{\can}{{\mbox{\footnotesize\bf c}}}

\newpage
\section {Introduction}

The  present paper  is devoted  to the  canonical
approach to quantum mechanics of the simplest quantum object,
a neutral spinless point--like particle,
in {\it the  general  Riemannian space--time} \ri with intention to compare
it in a subsequent paper with the version of the theory
resulting in the field--theoretical approach. The canonical
approach consists in quantization of the geodesical motion and
the field--theoretical one  starts with the quantum theory of the
real scalar field in \ri and, after  a specification of the
Fock space by a particle interpretation of states, restricts
naturally defined  second quantized operators of basic (primary)
observables   to  the  one--particle subspace.
Since in the general \ri there are no symmetry  arguments,
the particle interpretation can be based only on the concept
of the particle as a localized object, which is well developed
in the nonrelativistic quantum mechanics, see,  e.g. \cite{Jauch}, and
leads as a result of formalization of measurement procedures
to the Born's probabilistic  interpretation  of
the state vector and   position (coordinate) operators in the configuration
representation.

Motivation of the interest to  quantum  mechanics  in \ri is very
simple: consideration  of quantum systems in general  relativity
including quantization of gravitation itself would be incomplete
without study of consecutive generalization  of the basic principles
and notions of the standard quantum theory to the Riemannian space--time
background. This idea is supported, on my opinion,
by comparison of the two mentioned approaches which lead
generally to different versions of quantum mechanics in \ri
which, of course, are  consistent in the case of {\it the Minkowski
space--time} \eu.

 Quantization of geodesic motion in \ri
along  the mentioned above  standard lines has  not been  done
apparently in view of a widespread oppinion that actually it was considered
mathematically rigorously  by J.\'Sniatycki~\cite{Snia} in the framework
of the geometrical quantization.  However,
it is well--known that the same classical theory results in
different quantum counterparts depending on the procedure
of quantization chosen. In  \cite{Snia}  the  formalism
is chosen  which considers all space-time
coordinates as observables on an equal footing; I  discuss  it
very briefly  in Sec.2.  The  resulting  representation space in this
formalism is then $ L^2(\rif, {\bf C})$ and  thus
localization should be considered  on the space-time regions,
whereas the standard quantum--mechanical formalism is based in fact
on a 1+3--foliation of space-time and  on the concept of a particle
as a spatially localized object which is based on  formalization
of experimental procedures.

To this end,  in  Section 3 a   1+3--foliation
based on the normal geodesic extension of  a given Cauchy
hypersurface $\Sg$ is introduced and the classical  geodesic dynamics  is
represented  in the reduced Hamilton form
which is ready for canonical (i.e. \ri  is assumed to be  an
elementary manifold)  quantization.

In Section 4  quantization of
this system  is done  along the standard lines
 in the configuration space. It should be noted   that under
quantization  I mean  here  introduction of  primary (basic)
operators  of observables  of spatial position and  momentum
forming the so called quantum kinematics and construction
of the Hamilton operator in terms of the primary operators, which
provides with a  quantum dynamics. Thus, the problem of
construction of operators corresponding to any function on the phase
space (the full quantization) related to the problem of ordering of
the primary operators  is only designated.

As  a result, a  structure is obtained  which is very similar to the
ordinary non--relativistic quantum mechanics but it is, of course,
relativistic
in the sense that $c$, the velocity of light, is any finite parameter.
In Section 5 this structure is  represented
in a general covariant form which is  useful, in particular,
for further comparison with  quantum mechanics resulting
in the field--theoretical approach. In Section 6 a draft of
physical interpretation of the obtained  structure is exposed
on the basis of relation between idealized "yes--no" measurement
procedures forming a  quantum logic  and  projection  valued  measures
on  a  Hilbert space of states.

 It should be noted at once that  a heuristic (or naive) level of
mathematical rigor is adopted in  the paper and  a majority of  assertions
are of the   general situation, that is   necessary mathematical conditions
are supposed to be fulfilled.  In particular, a proper treatment
of unbounded operators, uniqueness and irreducibility
of introduced representations of observables are not considered at all.
I hope that it is plausible because my first aim is to outline possible
changes in quantum mechanics  related to introduction of the Riemannian
metric of  space--time.   A necessary mathematical refinement can apparently  be
made along the known lines of the standard theory if the
primary results and  further development of them prove to be interesting.

 Notation is standard  for general relativity
and, as a rule,  in the simple index form.
{\it The dot  between differential
operators denotes their operator product}, i.e. $\hat A \cdot \hat B$ means
that $\hat A \cdot \hat B\, \psi(x) \equiv \hat A (\hat B\psi (x))$.
Indices run the values: $\alpha,\, \beta,\, ... = 0,\, 1,\,2,\,3$ and
$i,\,j,\, ... 1,\,2,\,3$.

\section {Preliminary Notes on Quantization}

   Quantization, as it usually referred to P.A.M.Dirac~\cite{Dir},
is  a linear map
$ Q:  f \rightarrow \hat f $  of the Poisson algebra of  real functions
$ f \in C^\infty (M_{2n}) $  on a symplectic manifold $ (M_{2n},\, \Omega) $,
$\Omega$  being a symplectic form, to a set of operators acting
in a pre-Hilbert space  $\cal H$ ({\it the representation space}), provided
the following conditions are  fulfilled: \\

{\it 1)} $ 1\ \rightarrow {\bf\hat 1} $,  $\1$ being the unity operator in
$\cal H$ ;

{\it 2)} $ [f,\, g]_{\mbox{\footnotesize Poiss}}
\stc \Omega^{ab}\,\ptl_a f\,\ptl_b g \ \rightarrow \
i\h^{-1}   (\hat f \hat g - \hat g \hat f) \stc i\h^{-1} [\hat f, \hat g],
\quad  \ a,\, b,\,...\, = 1,\,...,\, 2n$ ;

{\it 3)} $ \ {\hat{\ov f}} = (\hat f)^\dagger $ on a dense subset of
$\cal H$, where {\it the dagger denotes the  Hermitean conjugation
with respect to the inner product in $\cal H$};

{\it 4)} a complete set  of commuting operators
$ \hat f_1, ..., \hat f_n $ exists, such that,  if exists $\hat f$
for which  $\left[\hat f,\ \hat f_i \right] = 0  $ for any  $ i $, then
$\hat f\ =\  \hat f (\hat f_1, ..., \hat f_n ) $.\\

 Unfortunately, these   conditions together are    apparently
intrinsically inconsistent  if one supposes arbitrary $C^\infty$ functions
$f$ and $g$  in the condition {\it 2)}   or (and) an arbitrary $M_{2n}$
in the condition {\it 4)}.  As concerns the condition {\it 2)},
one finds an analysis of it and numerous references  in \cite{kal}.
The solution of the inconsistency proposed there  is  that only canonical
coordinates $\{q,\,p\}$ in $M_{2n}$  should satisfy {\it 2)} whereas
commutators $[\hat f,\, \hat g]$  for other functions $f,\ g$ should be
determined  in addition. The present  paper is just restricted
to consideration of the primary observables $\{q,\,p\}$.

  As concerns {\it 4)}, the case of   dynamics of a point--like particle in
\ri  admits  a  mathematically  rigorous solution of  the problem  given
in \cite{Snia} where  $ M_8 \sim  T^* \rif $  is taken  the phase space.
 Canonical coordinates
$\{p_{(\alpha)}, \, q^{(\beta)} \}$ on $ T^* \rif $ are formed by a frame
$\{p_{(\alpha)}\}$ on the typical fiber  and  by values $ \{q^{(\beta)}\}$
of  four independent functions $ q^{(\beta)} (x),\ x \in \rif$ on the base
\ri. (I simplify the consideration assuming that \ri can be covered by a
single chart though essentially the geometrical quantization is
a method  for the  situations when this is not the case.)
Thus, $q^{(\beta)} (x)$ are classical observables defining a position
in \ri,  which  should be mapped on operators
by quantization,  whereas the coordinates ${x^\alpha}$ are defined
generally  on a chart $U \subset \rif$ and provide the space--time with
a primary manifold structure and are not quantized.

  In the introduced notation  the general--relativistic dynamics of a
particle of the rest mass  $ m $ on  \ri is determined by the constraint
\begin{equation}
 m^2 c^2 = p_{(\alpha)} p_{(\beta)}
\left(g^{\gamma\delta}\,
\ptl_\gamma q^{(\alpha)}\ptl_\delta q^{(\beta)}\right)\, (x) \circ \pi\ ,
\label{mc}  \\
\end{equation}
where $\pi$  is the projection in  $ T^* \rif  $,
i.e. any line in $ T^* \rif $ satisfying this condition is a time--like
geodesic.

   The result of consecutive quantization  of the described
structure can be presented, denoting operators in ${\cal H}$ by the "hat" 
mark, as follows~\cite{Snia}.
Let $ q(x) \in C^\infty (\rif) $ and $K_\alpha (x)$ is  a
$C^\infty (\rif)--\mbox{one--form (covariant vector field) over}\ \rif$
(a section in $T^* \rif $). Then
$$
Q:\ \{p_{(\alpha)}, \, q^{(\beta)} \}\ \rightarrow \
\{\hat p_{(\alpha)}, \, \hat q^{(\beta)} \}
$$
is equivalent to  the map
\begin{eqnarray}
q(x)\ & \rightarrow &\ \hat q(x) \stc   q(x)\cdot {\bf 1}, \label{q}\\
K_\alpha (x) \    & \rightarrow &\  \hat p_K (x)
\stc - i \h (K^\alpha \nabla_\alpha  + \frac12 \nabla_\alpha K^\alpha).
\label{p}
\end{eqnarray}
These operators act in $L^2 (\rif;\, {\bf C};\, \sqrt{-g} dx^0...dx^3)$,
i.e. in the space of complex functions $\varphi (x)$  with the inner product
\begin{equation}
 <\varphi_1, \varphi_2>
= \int_{\rif} \overline \varphi_1\, \varphi_2 \, \sqrt{-g} dx^0...dx^3.
\label{inn}
\end{equation}
The constraint Eq (\ref{mc}) is mapped by $Q $  to the condition
 specifying  in the representation space  a subspace of  functions
satisfying the  equation~\cite{Snia}
 \begin{eqnarray}
\Box\varphi + \zeta\, R(x)\, \varphi  + \left(\frac{mc}{\hbar}\right)^2
 \varphi &=& 0,  \quad x\in \rif \label{r} \\
\Box \stackrel{def}{=}  g^{\alpha\beta}\nabla_\alpha \nabla_\beta,
\qquad & &\nonumber
\end{eqnarray}
 with $\zeta = 1/6 $.  This is a very satisfactory result
because much earlier this   value of $\zeta$
was found to provide with the correct quasiclassical
behavior of one-particle states in the de Sitter--covariant
quantization of a scalar field \cite{cht}. From there immediately
follows the necessity of $\zeta = 1/6 $  in any \ri because there  is
no another scalar combination of components
of the Riemann--Christoffel tensor with  a dimensionless
coefficient, which provides with the same term for tha scalar field equation
in the de Sitter space--time \cite{tag}.

However, the structure so constructed is dissimilar to the
standard quantum mechanics with the Born probabilistic interpretation
in the configurational space,  where the representation space
is $L^2 (E_3;\, {\bf C};\,  dx^1...dx^3)$  and time  is
the evolution parameter,
not an observable.  If one would  consider $\vp (x)$ as a wave function of
a quantum object, this object is not stable because $\vp (x)$ have to
vanish in the time--like directions as well as in the space--like ones.
For example, any superpositions of the  usual  positive or negative
frequency plane wave solutions of Eq.(\ref{r}) in \eu   has an infinite
norm according to Eq. (\ref{inn}).
A more detailed consideration of this question  can be found in
\cite{tag1}.

It is clear that the roots of the divergence are in the choice of
 $ T^* \rif $ as initial  $M_{2n} $  and in the  symmetric  treatment of
space and time coordinates. Actually, a moment of time,  contrary to a
space position, is not a property  of a particle in the ordinary sense.
Quantization of the geodesical dynamics
 after some sort of the 1+3--foliation of \ri by   three--dimensional
configuration spaces enumerated by a time--like evolution parameter
would better correspond to this purpose.

\section {Classical Hamilton Theory  of Geodesics}

A time--like  geodesic in \ri is a line $x^\alpha =  x^\alpha (s) $
providing with an extremum of the action integral
between given points $x_1 = x(s_1),\  x_2 = x(s_2)$
\begin{equation}
W = - mc \int_{s_1}^{s_2}\, ds \sqrt{g_{\alpha\beta}
\frac{dx^\alpha}{ds} \frac{dx^\beta}{ds}}\ \stc \
\int_{s_1}^{s_2}\,L^\prime\,ds    \label{w}
\end{equation}
The Lagrangian $L^\prime$ is singular in the sense that the
components of four--momentum
\begin{equation}
p_\gamma (s) \stc \frac{d L^\prime}{d (dx^\alpha/ds)} =
- mc \, \left(g_{\alpha\beta}
\frac{dx^\alpha}{ds} \frac{dx^\beta}{ds}\right)^{-1/2}\, g_{\gamma\delta}\,
\frac{dx^\alpha}{ds}
\end{equation}
are not independent but  satisfy  the constraint
\begin{equation}
g^{\alpha\beta}p_\alpha p_\beta = m^2 c^2, \label{mcc}
\end{equation}
which is just Eq.(\ref{mc}) on a geodesic line
$\{p_{(\alpha)} = p_\alpha(s),\, q^{(\alpha)} = x^\alpha (s)\}$.
It is a consequence of invariance of the
action with respect to arbitrary "gauge" transformation $ s^\prime= f(s) $.
Owing to Eq.(\ref{mcc}),   the  Hamiltonian
\begin{equation}
H^\prime = p_\alpha\, \frac{dx^\alpha}{ds}\, - \, L^\prime
\end{equation}
vanishes: $H^\prime = 0$.

There exist methods of gaugeless quantization and they apparently
are equivalent to  "general--relativistic" quantization described
in Section 2. However just due to the arguments at the end
of the section,  I reduce  the Hamilton formalism
to the 1+3--foliated form
fixing a Cauchy  hypersurface $\Sg = \{ \Sg (x) = const,\ x\in \rif\}$,
introducing normal Gaussian coordinates $\{t, q^i\}$ based on $\Sg$,
in which
\begin{equation}
ds^2 = c^2\, dt^2 - \omega_{ij}(t, q) \,dq^i\,dq^j
\end{equation}
and  choosing   $ s = t $  in $W$:
\begin{equation}
W = - mc \int_{t_1}^{t_2}\, dt \sqrt{ c^2 -
\omega_{ij}(t, q) \,\dot q^i\,\dot q^j},
= \int_{t_1}^{t_2}\, dt \, L,          \label{w1}
\end{equation}
where $\dot q^i \stc d q^i/dt$. I suppose also that $t=0$ on the chosen
$\Sg$. Then  three independent space momenta are
\begin{equation}
p_i \stc \ \frac{\ptl L^\prime}{d\dot q^i}
= \frac{mc\ \omega_{ij}(t, q) \,\dot q^j}{\sqrt{ c^2 -
\omega_{ij}(t, q) \,\dot q^i\,\dot q^j}}.
\end{equation}
and the Hamiltonian is
\begin{equation}
H \stc p_i\,\dot q^i - L = mc \sqrt{c^2 -
\omega_{ij}(t, q^i) \,\dot q^i\,\dot q^j}.
\end{equation}
Now, the Hamiltonian form of geodesic equation is
\begin{equation}
\dot q^i = [H,\, q^i]_{\mbox{\footnotesize Poiss}},\  \
\dot p_i = [H,\, p_i]_{\mbox{\footnotesize Poiss}}. \label{haeq}
\end{equation}
The spatial coordinates $q^i (t) $  and momenta $p_i (t) $ are the basic
classical canonically conjugate (i.e. conjugate with respect
to the canonical symplectic form $\Omega$)  observables of a particle and
other classical observables are  real functions of them.

\section {Canonical Quantization of Geodesic Motion in \ri}

Now, let a representation space for quantum operators will be
$L^2 (\Sg;\, {\bf C};\, dq^1 dq^2 dq^3)$, i.e.  the space
of the square--integrable over $\Sg$ and sufficiently smooth
complex functions $\phi (q)$.  A Hilbert structure is defined
on it by  the inner product
\begin{equation}
<\phi_1,\,\phi_2>_\Sg \stc
= \int_\Sg \overline \phi_1\, \phi_2 \, dq^1 dq^2 dq^3, \qquad
\phi_1,\ \phi_2  \in  L^2 (\Sg;\, {\bf C};\, dq^1 dq^2 dq^3).\label{spr1}
\end{equation}
Introduce  operators $\check q^i \,,\ \check p_j $,  as follows
\begin{equation}
\check q^i   \stc   q^i \,\cdot\, \1,\quad
\check p_j  \stc - i \h\,\frac{\ptl}{\ptl q^j}    \label{qp}
\end{equation}
which    satisfy   the  Heisenberg commutation relations  and are
Hermitean (symmetrical) with respect to the inner product
Eq.(\ref{spr1}).  Thus,
they, together with the unity operator $\1$, satisfy     the conditions
{\it 1)} ---  {\it 3)} of quantization.
Then, take them as  operator Cauchy data on $\Sg$ for
the map $Q$ of the Hamilton equations Eq.(\ref{haeq}):
\begin{equation}
i\h \frac{d}{dt}\, \check q^i (t) =  [\check H (t),\, \check q^i (t)], \ \
i\h \frac{d}{dt} \check p_i(t)  = [\check H (t),\, \check p_i (t)]
\label{heeq}
\end{equation}
which are obviously the Heisenberg equations.  If
$\check H (t) \equiv \check H (t, \check q (t), \check p (t) )$
is a Hermitean  operator, then $\check q^i (t) \,,\ \check p_j (t)$
are also Hermitean operators owing to hermiticity of the Cauchy data.
Simultaneous hermiticity of $\check q^i (t) \,,\ \check p_j (t),$  and
$\check H (t)$ is provided by  the following  symmetric  substitution
of the basic operators  $\check q^i (t) \,,\check p_j (t)$
into the classical  Hamiltonian  $H (t)$:
\begin{equation}
\check H (t) = mc \sqrt{c^2 -
\check p_i (t)  \,\omega^{ij}(t, \check q(t))
 \,\check p_j (t)},  \label{ham2}
\end{equation}
However,  there is an infinite set of other possibilities of Hermitean
ordering of the cofactors $ \check p_i (t) \,,\ \check p_j (t)$
and $\omega^{ij}(t, \check q(t) $. This is  the well--known
ambiguity of ordering in construction  of a Hamilton
operator from a classical Hamiltonian. For definitness, I
take for further  consideration   Eq.(\ref{ham2}) as the Hamilton
operator; apparently this is the simplest choice.

Thus,  the conditions {\it 1)} ---  {\it 3)} of quantization are
satisfied  with regard to   the basic observables and the
Hamiltonian. As concerns
condition {\it 4)}, its consideration demands a formulation on
the level of geometric quantization that seems, in the
1+3--formalism , to be  much more  complicated than  in the
four--dimensional formalism mentioned in the Section.2 since one
should consider   different charts   on \ri and $\Sg$  simultaneously.
One should  also keep in mind the known problems of
mathematical justification of the use of  unbounded operators,
but, on the heuristic level accepted here, it is natural to suppose that
it can be done in a way similar to the  standard theory.

Now, let us  introduce  the Schr\"odinger wave functions
$\phi_{\mbox{\footnotesize Sch}} (t,\,q)$
in the usual way and, in addition,  transform them as follows:
\begin{equation}
\phi (q)\ \rightarrow \ \phi_{\mbox{\footnotesize Sch}} (t,\,q)
= \omega^{1/4}\,exp\left(-i \frac{mc^2}{\h}\, t\right)\,  \psi (t, q),
\quad  \omega \, \stc\,  {\rm det} \| \omega^{ij} (t, q)\|.
\end{equation}

The Heisenberg equations Eq.(\ref{heeq}) generate
  the following  Schr\"odinger  equation for $\psi (t, q)$
(the index $\can$ will mean further "canonical")
\begin{equation}
i\h\,(\frac{\ptl}{\ptl t} + \omega^{-1/4}\frac{\ptl\omega^{1/4}}{\ptl t})
\,\psi  = \hat H^\can (t, q, \hat p^\can) \, \psi.  \label{sche1}
\end{equation}
where
\begin{eqnarray}
\hat H^\can (t, q, \hat p^\can) & = &
 mc^2 \left( \sqrt{1 + \frac{2 \hat H^\can_0}{mc^2}} - 1\right)
\label{ham3}\\
\hat H^\can_0 &= &\hat H^\can_0 (t, q, \hat p^\can)
\stc \frac{1}{m}\, \hat p^\can_i \,\omega^{ij}(t, q) \,\hat p^\can_j
\end{eqnarray}
and
\begin{equation}
\hat p_i^\can \stc - i\h\,(\frac{\ptl}{\ptl q^i}
+ \omega^{-1/4}\frac{\ptl\omega^{1/4}}{\ptl q^i}).
\end{equation}
Note that $\hat H^\can $ is written as depending on the {\it c--numbers}
$q^i$ because, according to Eq.(\ref{qp}),   in the Schr\"odinger picture,
a coordinate operator is $\hat q^i_\can =  \check q^i\mid_\Sg \equiv
q^i \cdot \1 $.
It is easily seen also  that
\begin{equation}
\hat H^\can_0 = -\frac{\h^2}{2m}\, \triangle_t  +  V \label{ham0}
\end{equation}
where $\triangle_t$ is the Laplacian on  hypersurfaces
$$
t = \{\, \{t,\, q\} \in \rif; \  t = const; \ \Sg\, \sim\, t=0 \}
$$
enumerated by a value of the evolution parameter $t$; I  denote them
by the same letter $t$ since it does not cause a  confusion in a context.
\begin{equation}
V = V(t,\,q) =
\frac{\h^2}{32m} \omega^{ij} \ptl_i {\rm \log}\omega\,
\ptl_j {\rm \log}\omega
\end{equation}
is a so called {\it quantum potential}. Another  choice of ordering
in transition from $H$ to $\check H^\can_0$ would lead to a different
$V(t,\,q)$  but  the general structure of the Hamilton operator   retains
the form of Eq.(\ref{ham0}) for any ordering.  However, the question
of different quantum potentials deserves a further consideration.

Denote $\Psi$ the space of sufficiently smooth solutions of the \Sche
(\ref{sche1}). Then, from (\ref{spr1}) one has for
$\psi_1,\, \psi_2 \in  \Psi $ determined by  given intial
$\phi_1, \, \phi_2 \in L^2 (\Sg;\, {\bf C};\, dq^1 dq^2 dq^3)$
\begin{equation}
<\phi_1,\,\phi_2>_\Sg
= \int_\Sg \overline \psi_1(t, q)\, \psi_2 (t, q) \
\sqrt{\omega (t,q)}\,  dq^1 dq^2 dq^3 \stc (\psi_1,\, \psi_2)_t. \label{spr2}
\end{equation}
Thus,   $(\psi_1,\, \psi_2)_t$, which formally depends on $t$, actually
does not. It can be considered as a scalar product in $\Psi$
and written as
\begin{equation}
(\psi_1,\, \psi_2)_t = \int_t\,\overline \psi_1 \psi_2\ d\sigma_t
\label{spr3}
\end{equation}
where
$$d\sigma_t\,  = \,\sqrt{\omega (t,q)}\,  dq^1 dq^2 dq^3
= N_\alpha (x)\, d\sigma_t^\alpha (x) $$
is {\it the invariant volume element} of the hypersurface  $t$,
$\ N^\alpha (x)$ is {\it the normal geodesic extension} of the unit
normal  to $\Sg$, which is also a unit normal to $t$, containing
the point $x$, and $d\sigma_t^\alpha (x) $ is the normal volume
element of $t$.

However,   if $\hat O$ is an operator  which do not commute
with the Hamilton operator, then $\hat O\psi \not\in \Psi $  but as a
function on a given hypersurface $t$ it belongs to
$L^2 (t;\, {\bf C};\,d\sigma_t)$ with
the same scalar product $(\psi_1,\, \psi_2)_t$  the values of which
depend on $t$ if  $\psi_1,\, \psi_2 \not\in \Psi$  . Thus we have
an one--parametric family of $L^2$--spaces, each over a layer of
the introduced 1+3--foliation.

The (unbounded) operators $\hat q_\can,\ \hat p^\can $ and, owing to them,
the operator $\hat H^\can  $, are {\it Hermitean} in each
$L^2 (t;\, {\bf C};\,d\sigma_t)$, that  is
\begin{equation}
(\psi_1,\, \hat O  \psi_2 )_t = (\hat O \psi_1,\,   \psi_2 )_t,  \label{herm}
\end{equation}
on a  dense subset of $L^2 (t;\, {\bf C}; \,d\sigma_t)$.

\section {Covariantization}

Recall now that $t, q^i, $ are in fact functions of the   initially
introduced general coordinates $\{x^\alpha\}$ so that $ t = t(x), \
q^i= q^{(i)}(x)$ and
\begin{equation}
\ptl_\alpha t(x) \ptl^\alpha q^{(i)} (x)=0, \quad
{\rm rank}\,\|\ptl_\alpha q^{(i)} (x)\| = 3;  \label{qc}
\end{equation}
Enclosing the index $i$ in the brackets I denote that $q^{(i)} (x) $
as well as $t(x)$
are  {\it given functions, i.e. scalar functions with
respect to transformation of the coordinates} $x^\alpha$.
Thus, cf. Eq.(\ref{q}),
\begin{equation}
     \hat q_\can^i =\hat q_\can^{(i)} (x) = q^{(i)} (x) \cdot \1 .
\label{cqq}
\end{equation}

Introduce, further, three  vector fields $K_(i)^\alpha (x) $  on \ri
defined by  conditions
\begin{equation}
\ptl_\alpha t(x) K_{(i)}^\alpha (x) = 0,\quad
K_{(i)}^\alpha (x)  \ptl_\alpha\, q^{(j)} (x) = \delta_{(i)}^{(j)}
\label{pc}
\end{equation}
which  mean that $K_{(i)}^\alpha (x)$ lie on the hypersurface
$t_\Sg = \{x \in \rif;\  t(x) =const\}$
which contains the point $x$ and is conjugate to the coordinate $q^i$ on
$t_\Sg$.
Then the operator $\hat p_{(i)}$ can be represented as (cf. Eq.(\ref{p}))
\begin{eqnarray}
\hat p^\can_{(i)} =\hat p^\can_{(i)}(x)
&=& - i \h (K_{(i)}^\alpha \nabla_\alpha
+ \frac12 \nabla_\alpha K_{(i)}^\alpha)
\equiv \frac{i \h}{2} (K_{(i)}^\alpha\cdot \nabla_\alpha
+  \nabla_\alpha \cdot K_{(i)}^\alpha) \nonumber\\
&\equiv&- i \h (K_{(i)}^\alpha D_\alpha
+ \frac12 D_\alpha K_{(i)}^\alpha)
\equiv \frac{i \h}{2} (K_{(i)}^\alpha\cdot D_\alpha
+ D _\alpha \cdot K_{(i)}^\alpha) \label{p2}
\end{eqnarray}
where
\begin{equation}
D_\alpha \stc h^\beta_\alpha (x)\, \nabla_\beta \quad {\rm and} \quad
h^\beta_\alpha (x) \stc \delta^\beta_\alpha
- c^{-2}\ptl_\alpha t(x)\,\ptl^\beta t(x)
\end{equation}
is {\it the projection tensor on} $t$.
Thus, $D_\alpha$ is  {\it the projection of the covariant derivative on}
$ t$ and any observable
$\hat F (t, \hat q_\can , \hat p^\can) \equiv \hat F (t,  q, \hat p^\can)$
is an element of the closure  of Hermitean
differential operators of with coefficients depending on $x$, which
contain only the derivatives  $D_\alpha$ and are Hermitean in
$L^2 (t;\, {\bf C};\,d\sigma_t)$.

Further, let  ${\bf Q}$ and  ${\bf K}$  are the sets  of sufficiently
smooth functions $q (x)$  and  vector fields $K^\alpha (x)$
satisfying the first of conditions in Eq. (\ref{qc})
and Eq.(\ref{pc}) correspondingly. Since any $q(x) \in {\bf Q}$
and $K^\alpha (x) \in {\bf  K} $ can
be expressed correspondingly  as  a function of  given space position
functions $q^{(i)} (x)$ and as a linear combination of space basis
functions $K^\alpha (x)$, one can generalize the map
$$
Q_\can\, : \quad  q^{(i)} \rightarrow \ \hat q_\can^{(i)} (x),\quad
p_{(j)}\ \rightarrow \  \hat p^\can_{(j)} (x)
$$
to the map, cf. Eq.(2), (3),
\begin{eqnarray}
Q\,:\quad  {\bf Q} \ni q(x)\ \rightarrow \ \hat q_\can (x)\,
&=& q(x)\, \1,  \\
   {\bf K} \ni K^\alpha (x)\ \rightarrow \  \hat p^\can_K (x)
&=& - i \h (K^\alpha \nabla_\alpha + \frac12 \nabla_\alpha K^\alpha)
 \equiv \frac{i \h}{2} (K^\alpha\cdot \nabla_\alpha
+  \nabla_\alpha \cdot K^\alpha) \nonumber\\
 &\equiv& - i \h (K^\alpha D_\alpha +
\frac12 D_\alpha K^\alpha)
 \equiv \frac{i \h}{2} (K^\alpha\cdot D_\alpha
+  D_\alpha \cdot K^\alpha) \nonumber\\
\end{eqnarray}
These operators satisfy formally the generalized Heisenberg
commutation relations:
\begin{eqnarray}
 & [\hat q_\can,\, \hat q_\can^\prime ] = 0, &  \
\qquad q(x),\ q^\prime (x)\, \in\, {\bf Q} \label{qq} \\
 & [\hat p^\can_K,\, \hat p^\can_L]
= - i\h \hat p_{[K,\,L]_{\mbox{\footnotesize Lie}}}, &
\qquad K^\alpha (x),\ L^\alpha (x)\, \in\, {\bf K}, \label{kl}\\
 & [\hat q_\can, \hat p^\can_K ] = i\h K^\alpha \ptl_\alpha q,  &
\label{qk}
\end{eqnarray}
where $[K,\,L]_{\mbox{\footnotesize Lie}}$  is the Lie derivative
of a vector field $L^\alpha (x)$ along $K^\alpha (x)$,
that is $[K,\,L]_{\mbox{\footnotesize Lie}}^\alpha (x)
=  K^\beta\nabla_\beta L^\alpha - L^\beta\nabla_\beta K^\alpha$. These
relations were obtained in \cite[p.146]{Var}
>from a system of imprimitivity
for a symmetry  group $G$ acting on the configuration space
(i.e. $ t(x) =const$ in our case) so that ${\bf K}$ is the Lie
algebra of $G$. We came to these relations for the Lie algebra
of smooth vector fields on $t$ with no  assumption on
existence  of a symmetry   but  the mathematical rigor
of our consideration is essentially  less than that of \cite{Var}.

Thus, we have  constructed covariantly  what is often called
the kinematical structure of quantum mechanics of a particle and proceed
further with its dynamics.

The \Sche Eq.(\ref{sche1}) can  be rewritten now in the scalar (general
covariant) form as follows:
\begin{eqnarray}
i\h T(x) \,\psi(x)  & = & \hat H^\can (x, \hat p^\can)\,\psi(x)
\label{sche2}\\
T \stc c^2\,\left(\ptl^\alpha t(x) \ptl_\alpha +\frac12 \Box\, t(x)\right)
&\equiv& \frac{c^2}{2}
\left(\ptl^\alpha t(x)\cdot \ptl_\alpha 
+ \ptl_\alpha \cdot \ptl^\alpha t(x)\right)
\end{eqnarray}
where $\hat H^\can (x, \hat p^\can)$ has the same form of the right--hand
side of Eq.(\ref{ham3}) but with
\begin{equation}
\hat H^\can_0  = \hat H^\can_0 (x, \hat p^\can) \stc
\hat p^\can_i \, \ptl_\alpha q^{(i)} \ptl^\alpha q^{(j)} \,
\hat p^\can_j ,  \label{ham01}
\end{equation}
This nonrelativistic Hamilton operator has again the form of
Eq.(\ref{ham0}) with  the quantum potential
\begin{equation}
V = V (x) =
 \frac{\h^2}{8m} \ptl_\alpha K^\alpha_{(i)}\, \ptl_\gamma q^{(i)}
\ \ptl^\gamma q^{(j)}\, \ptl_\beta  K^\beta_{(i)}
\end{equation}
Thus, by means of the quantum potential,  the canonical Hamilton operator
essentially depends also on the choice of the independent functions
$q^{(i)} (x)$, which are in fact {\it the normal geodesic extensions} of the
functions on the initial Cauchy hypersurface $\Sg$
i.e. their values are constant along the geodesics which are normal
to $\Sg$.  It should be remarked that generally the
 normal geodesic extension  can be valid only in a limited normal
geodesic distance from $\Sg$ because, for any $\Sg$ except  the hyperplanes
in \eu,  the normal geodesics have intersections (focal points).
Can this complication be treated  by the Maslov-Fedoriuk methods
\cite{Mas}, is an interesting but  open question for me.

\section {A Draft of Interpretation}

In nonrelativistic quantum mechanics, relation between  physical experiments
over a quantum system and a the mathematical structure of quantum mechanics
is based generally on a correspondence between an idealized  "yes-no"
 experimental
device ("proposition")  and a $\sigma$--homomorphism  of  Borel
subsets $\Delta$ of a $\sigma$-algebra ${\cal S}$  to a lattice of subspaces
in a Hilbert space ${\cal H}$.  An excellent  presentation of this
topic is given by J.M.Jauch in the monograph \cite{Jauch} to which
I try to follow.

In our case of description of a particle in the configuration
space,  a Borel set $\Delta_t$  is generated on the  hypersurfaces $t$
 by intersections  of the latter with the normal geodesics that
issue from a Borel set $\Delta_\Sg$ on the initial hypersurface $\Sg$.
Thus,  $\Delta_t$ as a function of the parameter $t$  is
the normal geodesic extension of $\Delta_\Sg$.   One may also imagine it
as a Borel subset  over the set of all  geodesics normal to $\Sg$, but
with account of the remark at the end of the preceding section
and topological complications.

The  procedure of detection  of a particle in the Borel sets
$\Delta_t \subset \{t\,:\ t=const$  is associated usually with a  lattice in
${\cal H} \sim L^2 (t;\, {\bf C};\,d\sigma_t)$
generated  by  projection operators $\hat E(\Delta_t)$
defined as follows \cite[p.~1]{Jauch} :
\begin{equation}
 \left(\hat E(\Delta_t) \psi\right) (x) = \chi_{\Delta_t} \psi(x),
\quad \psi \in L^2 (t;\, {\bf C};\,d\sigma_t)         \label{e}
\end{equation}
where  $\chi_{\Delta_t} (x)$ is the characteristic function of
$\Delta_t$.
Thus one has a projection valued measure on
\begin{equation}
\Delta_t \rightarrow \hat E (\Delta_t)  \label{pvm}
\end{equation}
which determines  for each $\psi \in  L^2 (t;\, {\bf C};\,d\sigma_t) $
a numerical  measure
\begin{equation}
(\psi,\, \hat E (\Delta_t) \psi)_t  = \int_{\Delta_t} \,d\sigma_t \,
\ov\psi \,\psi
 \end{equation}
 which  can be interpreted, after normalization of $\psi$ to unity, as
a probability   to find a particle in $\Delta_t$
in the state determined  by $\psi$.
Of course, this is a refined formulation and generalization to \ri
of the well--known Born probabilistic interpretation of
function $\psi (x)$.
It is important to realize that we owe by this interpretation
to $L^2$--structure of ${\cal H}$ which is in fact a postulate
as well as the canonical commutative realization (\ref{e}) of system
of projections $\hat E (\Delta_t)$.

However, providing with the interpretation of vectors of the
representation space, the projection valued measure (\ref{pvm}) on $t$
defines nothing similar to  $\hat q^{(i)}$ corresponding  to a position
observable in the phase space of a particle.
In the case of   Cartesian coordinates $X^i$ in
 $\Sg \sim E_3$, the operators corresponding to a mesurment of them
are defined \cite{Jauch} through a
system of projections $\hat E \bigl(\Delta^i (\lambda_i)\bigr) $
associated with the class of subsets of $E_3$
\begin{equation}
\Delta^i (\lambda_i)
\stc \left\{X^1,\,X^2,\,X^3\,|\,X^i< \lambda_i\,\right\}, \quad
-\infty<\lambda_i<\infty,\quad i =1,\,2,\,3.              \label{l}
\end{equation}
This class  generates a $\sigma$--algebra and thus a Borel structure
on each line which is parallel to the  $X^{(i)}$--axis.

In our case of  a general $\Sg$ in \ri, having taken  three
functions $q^{(1)}(x),\, q^{(2)}(x),\,q^{(3)}(x),$ satisfying
conditions (\ref{qc}) one  may apparently generate
a $\sigma$--algebra ${\cal S}^{(i)}$  on each integral curve
of the vector field  of the $K_{(i)}^\alpha$ in $\Sg$ and in its normal
geodesic extension to $t$  by the  following  class of semiopen
subsets on  $\Sg$:
\begin{equation}
\Delta^{(i)} (\lambda_i)
\stc \left\{q^{1},\,q^{2},\,q^{3}\,|\,q^{(i)}< \lambda_i \,\right\}, \quad
 q^{(i)}_m \le \lambda_i < q^{(i)}_M, \quad i =1,\,2,\,3.       \label{li}
\end{equation}
where $q^{(i)}_m$ and $q^{(i)}_M$  are the minimal and maximal possible
values of  $q^{(i)}$. So it is taken into account that, in principle, the
coordinate curves  may have one or two endpoints or be closed
but one should have in mind possible  complications
in the curvilinear case  such as intersections
of different  coordinate lines of the same coordinate. A simple model example
is the intersection of  coordinate lines of $\phi $ in the center of
the spherical coordinates $r,\, \phi $ on  $E_2 $.

 Further, let  for definitness $i=1$   and let $\Delta^{(1)}$ is a  Borel set
of ${\cal S}^{(1)}$. One can  introduce a projection valued measure
\begin{equation}
{\cal S}^{(1)}\, \supset\, \Delta^{(1)} \ \rightarrow \
\hat E (\Delta^{(1)})    \label{dd}
\end{equation}
on each $q^{(1)}$--coordinate curve. Let us assume, in addition, that  it
is possible to introduce   projections $ \hat E (\Delta^{(1)}(\lambda_1))$
on the system of subsets Eq.(\ref{li}), which satisfy
the following  conditions (the spectral property):
\begin{equation}
\hat E \bigl(\Delta^{(1)}(\tilde \lambda_1)\bigr) \le
\hat E \bigl(\Delta^{(1)}(\lambda_1)\bigr)  \quad \forall\
\tilde \lambda_1 \le  \lambda_1 \quad {\rm and}\quad
\hat E \bigl(\Delta^{(1)} (q_m^{(1)}) \bigr)
= \emptyset,\ \hat E \bigl(\Delta^{(1)}(q_M^{(1)})\bigr) = {\bf 1}.
\end{equation}
Of course,  these conditions are not necessarily satisfied
in  any \ri and $\Sg$ but on our
heuristic level we can assume that the particle is  localised essentially
in a region where they take place.

   Then, one can introduce an operator $\hat q^{(1)} (x) $
 corresponding to measurement of  function $q^{(1)} (x)$   as follows.
At first, define all its  diagonal matrix elements as integrals
\begin{equation}
<\psi |\, \hat q^{(1)} (.)\, |\psi > \stc  \int_{q^{(1)}_m}^{q^{(1)}_M}
\, q^{(1)} (x) \ d \mu_\psi (\Delta^{(1)}) \quad {\rm}\ \forall\
|\psi>\, \in \,{\cal H} \label{dme}
\end{equation}
over a numeric measure
\begin{equation}
\mu_\psi (\Delta^{(1)}) \stc <\psi |\, \hat E (\Delta^{(1)})\, |\psi>
\end{equation}
related to  $|\psi>$,  which is generated by the pojection valued measure
(\ref{dd}) on each\\ $q^{(1)}$~--~coordinate curve.
Then, the   nondiagonal matrix elements are known through
the  polarization formula. Namely, any  real quadratic form
$A(\psi, \psi)$ defines a sesquilinear functional
$A^\prime(\psi_1,\, \psi_2)$, see, e.g., \cite[p.33]{Jauch}
\begin{equation}
A^\prime(\psi_1,\, \psi_2) =  A(\psi_1 + \psi_2) -
A(\psi_1 - \psi_2) + i A(\psi_1 - i\psi_2) - i A(\psi_1 + i  \psi_2)
\end{equation}

Again, in $L^2 (t;\, {\bf C};\,d\sigma_t)$, this construction
has  the canonical realization of the measure $\mu_\psi (\Delta^{(1)})$
through the characteristic function
$$
\chi (\Delta^{(1)})    = \left\{1 \qquad \mbox{for} \quad
q^{(1)} (x)\in \Delta^{(1)}, \atop
\  0 \qquad \mbox{for} \quad q^{(1)} (x)\not\in \Delta^{(1)}. \right.
$$
so that the integral Eq.(\ref{dme})  over this measure  reads
\begin{eqnarray}
(\psi,\, \hat q^{(1)}(.)\ \psi)_t
= \int _{q^{(1)}_m}^{q^{(1)}_M}\,  q^{(1)}(x)\
   d\int_{q^{(1)}_m}^{q^{(1)}_M }\,
\int_{q^{(2)}_m}^{q^{(2)}_M}\, \int_{q^{(3)}_m}^{q^{(2)}_M}\
dq^1\ dq^2\ dq^3
\,\sqrt{\omega_t}\ \chi (\Delta^{(1)}) \ \ov\psi \psi,
\qquad \nonumber \\
= \int _{q^{(1)}_m}^{q^{(1)}_M}\,
\int_{q^{(2)}_m}^{q^{(2)}_M}\, \int_{q^{(3)}_m}^{q^{(2)}_M}\
dq^1\ dq^2\ dq^3
\,\sqrt{\omega_t} \   \ov\psi\, q^{(1)}(x)\, \psi
 \quad \mbox {for any}\  \psi \in  L^2 (t;\, {\bf C};\,d\sigma_t)
 &&
\label{qqq}
\end{eqnarray}
Hence it is obvious  that, up to possible topological complications noted
above, Eq.(\ref{qqq}) defines the operator
$\hat q^{(1)}(x)$ just  as the multiplication operator
$\hat q_\can^{(i)} (x) $, Eq.(\ref{cqq}). Since the function $q^{(1)}(x)$
was chosen arbitrarily, the same is true for  any  $\hat q_\can (x)\,. $

I should emphasize  again that this reasoning is not a mathematically  rigor
justification of definition of the position operators but is presented
here for notification of possible (and necessary) development of
the theory, which seems also to be very interesting in the
nontrivial geometric context.

\section {Conclusion}

Thus, we introduced  a geodesic congruence (frame of reference) which
is normal to a given  Cauchy hypersurface $\Sg$ and,
by the simplest canonical procedure on this basis,
formulated basic notions and relations  of quantum mechanics
which corresponds to geodesic motion in the general \ri.
The resulting structure is similar to \Schp of the standard nonrelativistic
quantum mechanics with the Born probabilistic interpretation  of the
wave functions.

A specific feature of  the structure is that there are  two types
of coordinates. The first one consists of  $\{x^\alpha \}$, arbitrary
coordinates providing \ri with a  manifold structure. The second type
of coordinates are  $\{q^{(i)} (x)\} $  through which a  space localization
of a particle is determined and  the values of  which are assumed
to be outputs of specific devices, each  corresponding to  some
choice of a function $q^{(i)} (x)$.   These  coordinates
are  subjected to quantization and  are mutually commuting
quantum--mechanical operators. An attempt to generalize
the position operators to \ri is done though it
retains open questions of topological character which seem to be
very interesting for further study.

At the same time, the theory is relativistic  in the sense that
it is a quantum counterpart to a classical relativistic dynamics
and includes  $c$  as a parameter. If one takes \\
$ \hat p^\can_{(0)} (x) = mc^2 + \hat H^\can (x)$  then
\begin{equation}
 c^{-2}\,\left(\hat p^\can_{(0)}\right)^2\,
- \, \hat p^\can_{(i)} \,\omega^{ij}(t, q) \,\hat p^\can_{(j)} =
m^2 c^2 \cdot {\bf 1},
\end{equation}
that is just the relativistic relation for the four--momentum
of a particle of the mass $m$. One sees also  that the Hamilton operator
$H^\can $ is nonlocal. Computationally, it can be consdered asymptotically
in orders of $(mc^2)^{-1}$ and then the zero order (nonrelativistic) term
is, up to a choice of the quantum potential $V(x)$,
just the Hamiltonian which is  introduced  by a definition
in quantum mechanics on symmetric manifolds, see e.g. \cite{grosh} where
numerous references can be found.   Now the Hamilton operator is obtained
by quantization of a natural dynamics and one has a way to develop
study of symmetries to relativistic corrections if itis interesting.
Study of quantum potential in relation with the problem of
ordering of operators $\hat q^{(i)}, \ p_{(j)}$
seems  also  to be among the  important  questions arising in concern
with the present work. However, I should remind that my first aim is to
compare the present  canonical version of quantum mechanics with
that arising from the quantum field theory in \ri. It will be presented
in a subsequent paper II.

The author appreciates helpful discussions with Dr.D.Mladenov and
Dr. P.E. Zhidkov.

\end{document}